\documentclass[pre,twocolumn,groupedaddress]{revtex4-1}

\usepackage{graphicx,bbm,braket,color,lipsum,amsmath}

\begin{document}


\title{Impact of anisotropic interactions on non-equilibrium cluster growth at surfaces}


\author{Thomas Martynec}
\email[]{martynec@tu-berlin.de}
\author{Sabine~H.~L.~Klapp}
\email[]{klapp@physik.tu-berlin.de}
\affiliation{Institut f\"ur Theoretische Physik, Technische Universit\"at Berlin, Hardenbergstr. 36, 10623 Berlin, Germany}


\date{\today}

\begin{abstract}

Using event-driven kinetic Monte-Carlo simulations we investigate the early stage of non-equilibrium surface growth in a generic model with anisotropic interactions among the adsorbed particles. Specifically, we consider a two-dimensional lattice model of spherical particles where the interaction anisotropy is characterized by a control parameter $\eta$ measuring the ratio of interaction energy along the two lattice directions. The simplicity of the model allows us to study systematically the effect and interplay between $\eta$, the nearest-neighbor interaction energy $E_{n}$, and the flux rate $F$, on the shapes and the fractal dimension $D_{f}$ of clusters before coalescence. At finite particle flux $F$ we observe the emergence of rod-like and needle-shaped clusters whose aspect ratio $R$ depends on $\eta$, $E_{n}$ and $F$. In the regime of strong interaction anisotropy, the cluster aspect ratio shows power-law scaling as function of particle flux, $R \sim F^{- \alpha}$. Furthermore, the evolution of the cluster length and width also exhibit power-law scaling with universal growth exponents for all considered values of $F$. We identify a critical cluster length $L_{c}$ that marks a transition from one-dimensional to self-similar two-dimensional cluster growth. Moreover, we find that the cluster properties depend markedly on the critical cluster size $i^{*}$ of the isotropically interacting reference system ($\eta = 1$).
\end{abstract}


\maketitle

\section{Introduction}
The non-equilibrium surface growth of atomic systems by means of epitaxial layer growth has been intensively studied over the last decades. Several aspects from the sub-monolayer to the multilayer growth regime have been experimentally investigated in detail by atomic force microscopy \cite{exp_afm_1,exp_afm_2,exp_afm_3,exp_afm_4,exp_afm_5,exp_afm_6,exp_afm_7,exp_afm_8,exp_afm_9,exp_afm_10,exp_afm_11,exp_afm_12,exp_afm_13,exp_afm_14,exp_afm_15,exp_afm_16,exp_afm_17}, scanning tunneling microscopy \cite{exp_stm_1, exp_stm_2,exp_stm_3,exp_stm_4}, high or low energy electron diffraction \cite{exp_sem_1,exp_sem_2,exp_sem_3,exp_sem_4,exp_sem_5,exp_sem_6,exp_sem_7}, Raman \cite{exp_raman_1} and Auger electron spectroscopy \cite{exp_auger_1} experiments. Recently, also X-ray scattering studies of epitaxially grown thin films have been performed \cite{exp_xray_1,exp_xray_2,exp_xray_3,exp_xray_4,exp_xray_5,exp_xray_6,exp_xray_7,exp_xray_8}. Furthermore, such purely inorganic systems are theoretically well studied by means of rate equation approaches \cite{theo_rate_1,theo_rate_2,theo_rate_3,theo_rate_4,theo_rate_5,theo_rate_6,theo_rate_7,theo_rate_8,theo_rate_9,theo_rate_10,theo_rate_11,theo_rate_12,theo_rate_13,theo_rate_14,theo_rate_15} and kinetic Monte Carlo simulations (kMC) \cite{theo_kmc_1,theo_kmc_2,theo_kmc_3,theo_kmc_4}. Theoretical and numerical results for the cluster density, cluster size distribution, evolution of layer coverages and the global interface width are qualitatively in good agreement with experimental data for the growth of certain atomic systems. \cite{theo_rate_1, theo_rate_2, theo_rate_3, theo_rate_4, theo_rate_5, theo_rate_6, theo_rate_7, theo_rate_8, theo_rate_9, theo_rate_10, theo_rate_11, theo_rate_12, theo_rate_13, theo_rate_14}. Moreover, good agreement between experimental data and kMC simulations was also found for the growth of the organic molecule fullerence $C_{60}$ \cite{exp_xray_1}.

The above mentioned observables along with the shape of clusters in the sub-monolayer regime are of peculiar interest regarding the fact that clusters formed in the early stage of thin film growth provide the basis for further nucleation and growth in higher layers. Indeed, depending on properties of initially nucleated clusters, the morphology of the grown structure can drastically vary in the multilayer regime. This may strongly affect mechanical, optical and electrical properties of thin film devices \cite{exp_xray_3,exp_xray_8}.

In the present study we investigate the sub-monolayer growth of systems with anisotropic interactions. However, we note that even atomic systems can exhibit some kind of anisotropy. For example, Cu grown epitaxially on Pd(110) in the temperature regime below $300$ K has revealed a diffusion anisotropy, which is responsible for the formation of one-dimensional clusters \cite{aniso_atoms_1}. At higher temperatures, transverse diffusion of adsorbed Cu atoms sets in, leading to isotropic diffusion and the formation of regular two-dimensional clusters. Another example is the growth of Ag on fcc metal (110) surfaces: Quenched molecular-dynamics simulations \cite{md_aniso_atoms1,md_aniso_atoms2} have shown that the energy barriers and interaction energies for in-plane bonds parallel [$(1\bar{1}0)$] and normal [$(001)$] are not identical, which implies anisotropic interactions among adsorbed atoms. By varying the adsorption rate $F$ or the substrate temperature $T$, a rich variety of cluster morphologies from small isotropic clusters to one-dimensional and elongated two-dimensional clusters is observed \cite{aniso_atoms_2, aniso_atoms_3}. A further example involves the growth of Zn crystals on isotropic liquid surfaces. This system is known to produce rod-like and needle-shaped clusters with preferential lattice direction for particle attachment \cite{Lu2016}, and therefore also implies a form of interaction anisotropy among the adsorbed  Zn atoms. These studies indicate that the ratio $\Gamma = D_{0}(T)/F$ of the free diffusion $D_{0}(T)$ over adsorption rate $F$ not only influences the cluster density (which is well understood \cite{theo_rate_1, theo_rate_2, theo_rate_3}), but also the cluster morphology. However, details of the interplay between $\Gamma$ and cluster morphologies under non-equilibrium growth conditions in presence of interaction anisotropy are, so far, not well understood.

This is even more the case for systems of conjugated organic molecules (COM). In contrast to most atomic systems or systems of nearly spherical organic molecules like fullerene $C_{60}$, elongated organic molecules like diindenoperylene, p-sexyphenyl (6P), the perylene derivative PTCDI-C$_{8}$ or pentacene are known to generally interact anisotropically with each other when adsorbed on both, organic and inorganic substrates \cite{exp_xray_3,exp_xray_5,exp_xray_7,exp_xray_8,exp_aniso_1,theo_aniso_1,theo_aniso_2,theo_aniso_3,theo_aniso_4,theo_aniso_5}. Therefore one expects rather complex cluster shapes and corresponding changes in the cluster density, cluster size distribution and the coalescence behavior in the sub-monolayer growth regime as compared to atomic systems. But also when we tend towards multilayer growth, the behavior of organic and hybrid inorganic-organic systems (HIOS) can strongly differ from isotropically interacting systems. 

The structural and chemical flexibility of organic molecules is one of the main reasons for the production of hybrid inorganic-organic thin film devices. For example, the partial fluorinated derivative 6P-F$_{4}$ of the prototypical organic semiconductor $para-$sexiphenyl 6P is known to grow in a distinctly different morphology than 6P on the non-polar ZnO($10\bar{1}0$) surface \cite{exp_xray_3}. For 6P, needle-shaped clusters of flat lying molecules are found in the second layer. In contrast to this, fluorinated 6P-F$_{4}$ grows in an upright standing fashion with smoother surface morphology than 6P. These examples show the impact of small chemical variations, which change the anisotropic particle-particle and particle-substrate interactions, on the growth mode.

In order to get a deeper insight in how anisotropic interactions affect the non-equilibrium surface growth, we here study the sub-monolayer growth by means of event-driven kMC simulations \cite{ClarkeVvedensky98, Oliveira13} involving spherical particles with anisotropic nearest-neighbor interactions. The simulations are performed on a two-dimensional square lattice. Particles are adsorbed on the lattice at rate $F$ and hop between nearest-neighbor sites until they meet other particles, yielding in-plane bonds that reduce the hopping rate. We vary the interaction energy and the degree of anisotropy of bonds to study the effect of modified interparticle interactions on structurally altered organic molecules to mimic effects like different polarities. Thereby, the effect of interaction anisotropy on the shapes of clusters formed in the very early stage of thin film growth is analyzed in detail. 

The rest of the paper is organized as follows. In Sec. II A, we describe the event-driven kMC simulation setup and the growth model with anisotropic interactions. In addition, we introduce in Sec. II B an anisotropic version of the Eden growth model. Numerical results for cluster properties for different system settings under non-equilibrium growth conditions are presented in Sec. III. We close with a brief summary and conclusions in Sec. IV.

\section{Model and Methods}

\subsection{The kMC model with anisotropic interactions}

During the non-equilibrium growth process simulated by a kMC algorithm, particles are adsorbed on an initially empty square lattice with an effective adsorption rate $F$ given in monolayer per minute (ML/min). Once adsorbed, they perform activated Arrhenius-type hopping processes to a randomly chosen nearest-neighbor lattice site. The hopping rate $r_{ij} \sim \text{exp}(-\beta \Delta E)$ from lattice site $i$ to a neighboring site $j$ is determined by an activation energy barrier $\Delta E$ which involves up to three contributions: (I) an in-plane diffusion barrier $E_{\text{d}}$, (II) an additional out-of-plane diffusion barrier for hopping across step-edges $E_{\text{es}}$, and (III) a nearest-neighbor interaction energy contribution $E_{\text{n}}$. In systems with isotropic nearest-neighbor interactions, the corresponding energy contribution (III) depends on the interaction energy $E_{n}$ of a two particle bond and on the number $n = \sum_{\braket{ij}}o_{ij}$ of occupied in-plane nearest-neighbor lattice sites (where $o_{ij} = 0$ if the neighboring site $j$ is unoccupied and $o_{ij} = 1$ if $j$ is occupied). The total contribution of the interaction energy to the hopping rate $r_{ij}$ then reads $\sum_{\braket{ij}} o_{ij} E_{n} = n E_{n}$. Here we consider anisotropic nearest-neighbor interactions, where not only the number $n$ of in-plane bonds, but also their configuration matters.

To this end, we define the interaction anisotropy parameter $\eta \in [0,1]$ which changes the nearest-neighbor interaction energy of in-plane bonds along the y-axis ($E_{n}^{y} = \eta E_{n}$) relative to that along the x-axis ($E_{n}^{x} = E_{n}$). Thereby, we model generic properties of anisotropic interactions (which are essentially omnipresent for conjugated organic molecules, but also for some atomic systems \cite{aniso_atoms_2, aniso_atoms_3,Lu2016}) combined with a global symmetry breaking e.g., an external electric field. 

Possible real systems corresponding to our model might be organic oligomers on the ($10\overline{1}0$) surface of a ZnO semiconductor. The surface generates an electric field that induces dipole moments in the adsorbed molecules along the field direction \cite{DellaSala2011}, yielding anisotropic dipolar interactions. To give a further example, anisotropic dipolar interactions occur between partially fluorinated organic molecule such as di-fluorinated para-sexiphenyl (6P-F2) \cite{exp_xray_3}. In our model, setting $\eta = - 1/8$ mimics the interaction of parallel aligned neighboring dipoles. Furthermore, the situation with $\eta < 1$ may describe systems where a lattice direction of preferred particle attachment exists. This is the case for the growth of elongated and needle-shaped Zn crystals on isotropic surfaces \cite{Lu2016} and the growth of Ag clusters on fcc metal (110) surfaces \cite{aniso_atoms_2, aniso_atoms_3}. In principal, both of these systems can be considered with our model.

One important peculiarity of our system is the fact that the particle shape remains isotropic. This allows us to study the impact of anisotropy in the interactions alone, without accounting for steric effects. Clearly, the latter effects are ubiquitous in a lot of realistic anisotropic systems such as films of organic molecules. However, from the simulation perspective, anisotropic particle shapes lead to additional complications such as blocked pathways for hopping processes, overhangs of adsorbed particles and the difficult question how the out-of-plane diffusion of anisotropically shaped particles should be treated \cite{theo_aniso_4,Goose10,Hlawacek08,Klopotek_I,Klopotek_II}. We consider the present simplified model as a first step to the overall goal to better understand the effect of anisotropic interparticle interactions under non-equilibrium growth conditions.

An illustration of the kMC model setup is shown in Fig. 1. The total interaction energy of a particle at site $i$ reads $(n_{x} + \eta n_{y}) E_{n}$, where $n_{x}$ is the number of occupied lateral neighbor sites along the x-direction, while $n_{y}$ is the same for the y-direction. For $\eta = 1$, the model thus reduces to the isotropic case [with total interaction energy $(n_{x} + n_{y}) E_{n} = n E_{n}$, while $\eta < 1$ represents the situation with anisotropic interactions among the adsorbed particles. Specifically, decreasing $\eta$ leads to an increase of the anisotropy of interparticle interactions. The resulting expression for the hopping rate from an initially occupied site $i$ to the final site $j$ is given by the Clarke-Vvedensky bond-counting Ansatz \cite{ClarkeVvedensky98, Oliveira13}, 

\begin{equation}
r_{ij} = \nu_{0} \, \text{exp}\{-\beta \left[(n_{x} + \eta n_{y}) E_{n} +  E_{\text{d}} + s_{ij} E_{\text{es}}\right]\}.
\end{equation}

Here, we have introduced the attempt frequency $\nu_{0} = 2k_{B}T/h$, where $k_{B}$ is the Boltzmann's constant, $T$ the substrate temperature and $h$ the Planck constant. Further, $\beta = 1/k_{B}T$ The first term in the exponent [$(n_{x} + \eta n_{y})E_{\text{n}}$] describes the contribution to the total activation energy barrier $\Delta E$ that stems from in-plane interparticle bonds, while the second and third term represent the in-plane ($E_{\text{d}}$) and out-of-plane diffusion barrier ($E_{\text{es}}$), respectively. The latter leads to a reduced rate for diffusion processes across step-edges (where $s_{ij} = 1$) by a factor $\alpha = \text{exp}(-\beta E_{\text{es}})$. The case $s_{ij} = 0$ corresponds to pure in-plane diffusion.

The simulation consists of a sequence of iterations. After each iteration step, where either a particle performed a hopping process to a randomly chosen nearest-neighbor lattice site or a new particle got adsorbed, the simulation time is updated in a stochastic manner by a time step $\tau$. The latter is calculated according to

\begin{equation}
\tau = -\frac{\text{ln}(R)}{r_{\text{all}}},
\end{equation}

where $R \in \left( 0,1 \right]$ is a random number which is chosen uniformly from the given interval and $r_{\text{all}}$ is the sum of rates related to all particles adsorbed in the topmost layer on the surface. In other words,

\begin{equation}
r_{\text{all}} = \sum\limits_{i = 1}^{L^{2}} \left(\sum\limits_{j = 1}^{4}r_{ij} + F\right),
\end{equation}

where $L$ is the lateral length of the discretized simulation box. Following earlier studies \cite{theo_kmc_1,theo_kmc_2,theo_kmc_3,theo_kmc_4}, we do not allow the collective diffusion of clusters and forbid overhangs and vacancies in the simulations. This means that the solid-on-solid (SOS) condition is applied to the system.

For all simulations in this work, we set the energy barrier for free diffusion to $E_{\text{d}} = 0.5$ eV. We choose this value because it is a good approximation for several real systems ranging from organic molecules like $C_{60}$ to inorganic systems like Ag or Pt \cite{exp_xray_1, theo_kmc_1}. The out-of-plane diffusion barrier is set to $E_{\text{es}} = 0.1$ eV for the same reason. The adsorption rate is varied between $F = 1$ ML/min and $F = 100$ ML/min, while the temperature is fixed to $T = 313$ K, a commonly used temperature in experimental growth studies with organic molecules \cite{exp_xray_1, Khokhar2012}. If not stated otherwise, the coverage is set to $\theta = 0.05$. This low coverage is chosen to make sure that coalescence of clusters has not yet set in. The simulations are performed at different values of the interaction energy and the anisotropy parameter in the ranges $E_{\text{n}} \in [0.10 - 3.0]$ eV and $\eta \in [0,1]$, respectively. By this we study the interplay between $E_{\text{n}}$ and $\eta$ concerning properties of growing clusters in the sub-monolayer growth regime when anisotropic interactions are present.

\begin{figure}
	\includegraphics[width=1.0\linewidth]{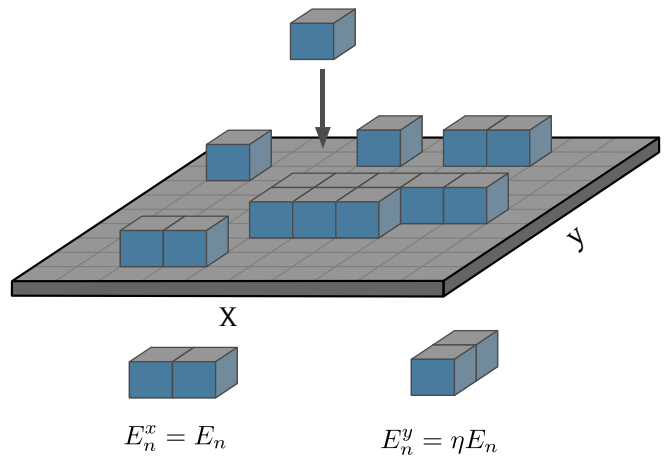}%
	\caption{\label{Average cluster shape} (Color online) Illustration of our model system for the non-equilibrium cluster formation and lateral growth in presence of anisotropic interactions. The in-plane interaction between particles on nearest-neighbor sites in x-direction is denoted by $E_{n}^{x} = E_{n}$. The parameter $\eta$ controls the degree of interaction anisotropy. For $\eta = 1$ interactions are isotropic, while for $\eta < 1$, the interaction energy $E_{n}^{y} = \eta E_{n}$ of bonds in y-direction is lowered relative to  $E_{n}^{x}$.}
\end{figure}

\subsection{Anisotropic Eden growth model}

The kinetic Monte-Carlo algorithm described in Sec. II. A. mimics the kinetically driven growth of thin films based on Arrhenius-type activation energy-dependent process rates. One goal of the present study is to compare the kMC results with those from an anisotropic stochastic Eden growth model. The latter is more elementary in the sense that it simulates cluster growth simply by attachment of particles to an existing cluster. This implies essentially the neglection of computationally costly hopping processes that usually dominate in kMC simulations, especially under realistic growth conditions \cite{eden_1,eden_2,eden_3,eden_4,eden_5,eden_6,eden_7}.

Within the Eden model, a cluster on the discretized two dimensional lattice space $\mathbbm{L}^{2}$ is defined as a finite subset $C \in \mathbbm{L}^{2}$ of occupied lattice sites. At the boundary of such a cluster, unoccupied sites $\partial C$ that possess at least one occupied neighbor site

\begin{equation}
\partial C = \{ j \in \mathbbm{Z}^{2} \ C: \exists i \in C \rightarrow \Vert i-j \| = 1 \},
\end{equation}

represent the set of growth sites which have a non-zero probability to be occupied in an iteration step during the cluster growth process. On a square lattice there exist four different types of nodes $\partial_{k} C$ ($k$ = 1,2,3,4), where $k$ is the number of occupied neighbor sites. Therefore, the total boundary is simply given by $\partial C = \partial_{1} C \cup \partial_{2} C \cup \partial_{3} C \cup \partial_{4} C$. 

The cluster at initial time $t_{0} = 0$ is a fixed connected set $C_{0} \subset \mathbbm{Z}^{2}$. In our case, the initial cluster at $t_{0}$ consists of just a single occupied site in the middle of the lattice. Thus, there exist four growth sites in the first iteration step. In each step $t_{n} \rightarrow t_{n+1}$ one of the growth sites $\partial C$ is occupied and the cluster grows $C_{n} \rightarrow C_{n+1}$ by one lattice site. The probability $p_{i}$ for particle attachment at boundary site $i$ depends on it's local environment, namely the number of occupied nearest-neighbor sites $s^{nn}$

\begin{equation}
p_{i} = \sum_{\braket{i,j}} s_{j}^{nn},
\end{equation}

where $s_{j}^{nn} = 1$ if the neighboring site is occupied, $s_{j}^{nn} = 0$ for unoccupied neighbor sites, and the sum $\braket{i,j}$ is taken over all nearest-neighbors of site $i$. Therefore, only growth sites with at least one neighboring cluster site have a non-zero probability $p_{i}$ to be occupied during the cluster growth process. 

In order to model anisotropic interactions, we split the occupation probability $p_{i}$ for lattice site $i$ into two contributions related to the x- and y-direction and impose an imbalance between these two directions. The anisotropy parameter $\xi$ determines the reduced attachment probability for particles along the y-direction \cite{eden_3}

\begin{equation}
p_{i} = p_{i}^{x} + p_{i}^{y} = \sum_{\braket{i,j}^{x}}s_{j}^{nn} + \xi \sum_{\braket{i,j}^{y}} s_{j}^{nn}.
\end{equation}

Here, $\braket{i,j}^{\text{x}}$ denotes occupied neighbors along the x-direction and $\braket{i,j}^{\text{y}}$ along the y-direction. Consequently, $p_{i}^{\text{y}} < p_{i}^{\text{x}}$, if $\xi < 1$. We normalize all probabilities $\tilde{p_{i}} = p_{i} / \sum_{j=1}^{c} p_{j}$ such that $\sum \tilde{p_{i}} = 1$. The attachment anisotropy in the stochastic Eden model mimics the interaction anisotropy in the kMC simulations and it results in the formation of elongated clusters for $\xi < 1$ \cite{eden_3}.

The actual simulation proceeds as follows. In each iteration step, we pick a random number $r \in [0,1]$ from a uniform interval. If $r \in [\tilde{p}_{i},\tilde{p}_{i+1}]$ we choose site $i$ to be occupied in this step. Different values for $\xi$ are used to study the effect of anisotropic interactions on the growth of clusters. The results are compared to clusters from the kMC simulations to check whether this minimal model is able to produce clusters with the same properties.
 
\subsection{Target quantities}

\subsubsection{Spatial extension of clusters}

In order to study how anisotropy in the interparticle interaction affects the shape of growing clusters, we calculate the average length $L$ of clusters in x-direction as function of cluster size $S$. The latter is the number of particles a cluster consists of. We further calculate the extension of the cluster in y-direction, i.e. the cluster width $W$. This also yields the aspect ratio $R = L/W$ which we calculate for different cluster sizes $S$ as function of interaction energy $E_{n} \in [0.15,1.0]$ eV and anisotropy parameter $\eta \in [0,1]$. The obtained results are averaged over at least 1000 clusters for each cluster size $S$. Furthermore, we calculate the cluster size distribution $P(S)$ and the distribution of cluster lengths $P(L)$ in order to analyze not only average quantities, but also fluctuations around the average.

\subsubsection{Fractal dimension of clusters}

An additional measure of the cluster morphology is the mass fractal dimension $D_{f}$ that describes the scaling of the cluster size $S$ (or mass) with the radius of gyration via \cite{Heinson2010, Sorensen1997, Schaefer1984, Stepto2015, Rahbari2012}
	
\begin{equation}
S = k_{0} (R_{g} / a)^{D_{f}}. 
\end{equation}
	
In Eq. (7), the value of the constant pre-factor $k_{0}$, depends on the cluster shape \cite{Heinson2010, Sorensen1997} and is of order unity. Further, $R_{g}$ is the radius of gyration and $a$ represents the particle radius which we set to $1$. We determine $D_{f}$ (as function of interaction energy $E_{n}$ and anisotropy parameter $\eta$) via the inertia tensor which, for a cluster in the x-y plane consisting of $S$ particles, is given by \cite{Heinson2010, Vymetal2011, Theodorou1985}

\begin{equation}
\textbf{T} = \sum_{i=1}^{S}
\begin{pmatrix} y_{i}^{2} & -x_{i} y_{i} & 0 \\ -x_{i} y_{i} & x_{i}^{2} & 0 \\ 0 & 0 & x_{i}^{2} + y_{i}^{2} \end{pmatrix}
\end{equation}

The eigenvalues $E_{i}$ ($i= 1,2,3$) of $\textbf{T}$, sorted according to their size ($E_{1} \ge E_{2} \ge E_{3}$), define the square of the principal radii of gyration via $R_{i}^{2} = E_{i} / S$, with $S$ being the cluster size. Thus,  $R_{1} \ge R_{2} \ge R_{3}$. From the quantities $R_{i}$, the radius of gyration (given in lattice sites) follows as

\begin{equation}
R_{g} = \sqrt{\frac{1}{2} (R_{1}^{2} + R_{2}^{2} + R_{3}^{2})}.
\end{equation}

The precise value of the cluster shape-dependent pre-factor $k_{0}$, that accounts for the cluster shape anisotropy (elongation of the cluster), depends on the ratio of the largest over the smallest squares of principle radii of gyration, that is, $A_{13} = R_{1}^{2} / R_{3}^{2}$ \cite{Heinson2010, Fry2004_1,Fry2004_2}.

\subsubsection{Cluster density}

A further important quantity is the cluster density $\rho_{N} = N / L^{2}$ (where $N$ is the number of "stable`` clusters in the system of size $L^{2}$) for different values of the interaction energy $E_{n}$ and the anisotropy parameter $\eta$. Clusters are considered as stable when they grow during the simulation time by subsequent particle attachment rather than dissolve into individual particles again. The latter are called unstable clusters. The distinction between these two types of clusters is typically associated with the critical cluster size $i^{*}$ which is defined as the largest size of unstable clusters such that clusters of size $S \le i^{*}$ dissolve while clusters of size $S > i^{*}$ grow. In the early stage of growth, the number of stable clusters, and thus, the cluster density $\rho_{N}$ increases until it saturates at $\rho_{N}^{c}$. This maximum value of the cluster density is referred to as the critical cluster density $\rho_{N}^{c}$. According to a rate equation approach \cite{theo_rate_12,theo_scaling_1}, $\rho_{N}^{c}$ is connected to the critical cluster size $i^{*}$ via the scaling relation $\rho_{N}^{c} \sim (D_{0}(T)/F)^{-\chi}$, with the exponent $\chi = i^{*}/(i^{*}+2)$, the rate for free in-plane diffusion $D_{0}(T) = \nu_{0} \text{exp}(-\beta E_{\text{d}})$ and the adsorption rate $F$. We are particularly interested in the critical cluster size of the systems with isotropic interactions ($\eta = 1$) for different interaction energies $E_{n}$. We take these isotropic systems as ``reference" systems because, as will be shown in Sec. III C, the precise value of $i^{*}$ in these systems has strong impact on the cluster shape properties in presence of anisotropic interactions when the nearest-neighbor bond strength  in y-direction is reduced to $E_{n}^{y} = \eta E_{n}$ ($\eta < 1$).

\subsubsection{The critical cluster size, reversible versus irreversible cluster growth}

The general procedure to determine $i^{*}$ for the isotropically interacting reference systems is as follows. First, $\rho_{N}^{c}$ is calculated for different $\Gamma = D_{0}(T)/F$ by changing the temperature $T$. We then choose $i^{*}$ such that it fits the numerically obtained scaling of the critical cluster density $\rho_{N}^{c}$. For $i^{*} = 1$ $(\chi = 1/3)$, already dimers represent stable clusters which do not decay. In this case, the critical cluster density scales as $\rho_{N}^{c} \sim \Gamma^{-1/3}$. In other words, the case $i^{*} = 1$ corresponds to \textit{irreversible} attachment, where particles become immediately immobilized for the rest of the growth procedure once they form at least one in-plane bond to a neighboring particle. In contrast to this, for $i^{*} > 1$ particle attachment is \textit{reversible} in the sense that they may detach from clusters, diffuse further and attach again to the same or some other cluster in the system. In this case $\chi < 1/3$, and consequently $\rho_{N}^{c}$ scales differently compared to the situation $i^{*} = 1$. A major problem with this procedure is that we need to perform simulations at different $\Gamma$ to obtain the best fit for the scaling of $\rho_{N}^{c}$ that determines $i^{*}$ when $T$, or $F$ are varied. However, often we are interested only in the question whether cluster growth is \textit{reversible} or \textit{irreversible} at a specific value of $\Gamma$. 

As an estimate, we calculate the sum of all hopping events at all time steps $t_{0},t_{1},t_{2},.., t_{f-1},t_{f}$  [with $t_{n+1} = t_{n} + \tau$, see Eq. (2)] of the simulation. In this time series $t_{0} = 0$ corresponds to the empty lattice at the start of the simulation, while $t_{f}$ corresponds to the time at the end of the simulation when the final coverage of $\theta = 0.05$ has been reached. In each iteration step $t_{n} \rightarrow t_{n+1}$, where a particle performs a hopping event, we distinguish between free hopping without lateral neighbors ($n = 0$) corresponding to a rate $r_{ij} \sim \text{exp} (-\beta E_{d})$ (denoted as $r^{0}$) and hopping processes of particles with at least one in-plane bond ($n > 0$) with rate $r_{ij} \sim \text{exp} (-\beta E_{d} + n E_{n})$ (denoted as $r^{>0}$). We then calculate the sum $R_{0}$ of all hopping events with $r^{0}$ and the sum $R_{> 0}$ of all events with $r^{>0}$ during the entire simulation, that is,

\begin{equation}
R_{0} = \sum_{t = t_{0}}^{t_{f}} r^{0}(t), \qquad r^{0}(t) =
\begin{cases}
1, & n_{t} = 0 \\
0, & n_{t} > 0
\end{cases}
\end{equation}

\begin{equation}
R_{> 0} = \sum_{t = t_{0}}^{t_{f}} r^{>0}(t), \qquad r^{>0}(t) =
\begin{cases}
1, & n_{t} > 0 \\
0, & n_{t} = 0.
\end{cases}
\end{equation}

Here, $n_{t}$ is the number of in-plane bonds of the particle that performed a hopping process at time $t_{n}$. From these two sums, we calculate the quantity 

\begin{equation}
\omega_{1} = \frac{R_{0}}{R_{> 0}},
\end{equation}

measuring the ratio between free hopping events and hopping events with in-plane bonds. If cluster growth is \textit{irreversible} ($i^{*} = 1$), no detachment events occur. Thus, $R_{> 0} = 0$ and consequently $\omega_{1} = 0$. In contrast, non-zero values of $\omega_{1}$ ($R_{> 0} > 0$) indicate the presence of detachment events or, in turn, \textit{reversible} cluster growth ($i^{*} > 1$). This method does not allow a precise determination of the exact of value of $i^{*}$, but it is sufficient to distinguish between the cases $i^{*} = 1$ and $i^{*} > 1$. We will use the quantity $\omega_{1}$ in Sec. III. C to differentiate between \textit{irreversible} or \textit{reversible} cluster growth conditions in the isotropically interacting reference systems.

\section{Results}

In the following, we present numerical results of kMC simulations for the non-equilibrium surface growth with interaction energies ranging from $E_{n} = 0.10$ eV to $E_{n} = 3.0$ eV and interaction anisotropies $\eta \in [0,1]$. First, we mainly focus on two exemplary interaction energies and values of the anisotropy parameter. These are $E_{n} = 0.2$ eV and $E_{n} = 0.7$ eV and $\eta = 1$ (isotropic interactions) as well as $\eta = 0.1$ (strongly anisotropic interactions). We chose these two values of $E_{n}$ because, for isotropic interactions, $E_{n} = 0.7$ eV represents a reference system with $i^{*} = 1$, while at $E_{n} = 0.2$ eV, we find $i^{*} = 2$ from scaling of the critical cluster density $\rho_{N}^{c}$. Therefore, for $\eta = 1$ these two values of $E_{n}$ represent growth conditions with \textit{reversible} and particle \textit{irreversible} attachment, respectively. 

\subsection{Spatial configurations and cluster shapes}

\begin{figure}
	\includegraphics[width=1.0\linewidth]{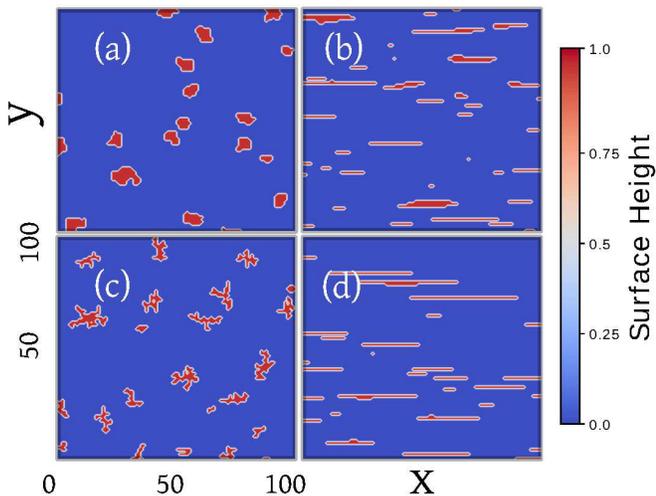}%
	\caption{\label{Average cluster shape} (Color online) Spatial configurations at coverage $\theta = 0.05$, $T = 313$ K, $F = 1$ ML/min (used in all figures). In (a) and (b) the interaction energy is $E_{n} = 0.2$ eV. In (a) the interactions are isotropic ($\eta = 1$) while in (b) strong anisotropy in the inter-particle interaction is present ($\eta = 0.1$). The configurations in (c) and (d) depict the same, but at higher interaction energy $E_{n} = 0.7$ eV.}
\end{figure}

\subsubsection{Fractal Dimension, spatial configurations and distributions}

As a starting point, we present snapshots of spatial configurations of surface structures in the sub-monolayer growth regime at coverage $\theta = 0.05$ ML in Fig. 2 for two interaction energies, $E_{n} = 0.2$ [Fig. 2(a) and (b)] and $E_{n} = 0.7$ [see Fig. 2(c) and (d)] eV for isotropic ($\eta = 1$) and strongly anisotropic ($\eta = 0.1$) growth conditions at $T = 313$ K and $F = 1$ ML/min.

For isotropic growth conditions ($\eta = 1$) the clusters have compact shapes at $E_{n} = 0.2$ eV [see Fig. 2(a)], while they are strongly ramified for the much stronger interaction energy $E_{n} = 0.7$ eV [see Fig. 2(c)]. This can be explained via corresponding values of the critical cluster size $i^{*}$. At $E_{n} = 0.2$ eV, we find $i^{*} = 2$ ($\chi = 1/2$). This implies \textit{reversible} attachment, i.e., particles with only one lateral bond may detach from clusters in order to attach to cluster boundary sites with a higher coordination number (i.e., number of occupied in-plane nearest-neighbor lattice sites). These particle rearrangements lead to compact clusters with fractal dimension $D_{f} \approx 2$. This is shown in Fig. 3 where $D_{f}$ versus $\eta$ is plotted for different values of $E_{n}$. We observe that, at $E_{n} = 0.2$ eV, $D_{f}$ is not affected by anisotropic interactions and remains close to $D_{f} \approx 2$ for all $\eta \le 1$. At $E_{n} = 0.7$ eV, attachment of particles to clusters is \textit{irreversible} ($i^{*} = 1$, $\chi = 1/3$). This leads to ramified cluster shapes [see Fig. 2(c)] where already one lateral bond is strong enough to suppress particle detachment. Specifically, we find that, at isotropic growth conditions, the fractal dimension in this case is $D_{f} \approx 1.7$ (also for $E_{n} = 0.3$ eV and $E_{n} = 0.5$ eV). The latter value is close to $D_{f} = 1.71$, the fractal dimension of clusters grown by diffusion-limited aggregation (DLA) \cite{dla_1,dla_2,dla_3,dla_4,dla_5}. 

\begin{figure}
	\includegraphics[width=1.0\linewidth]{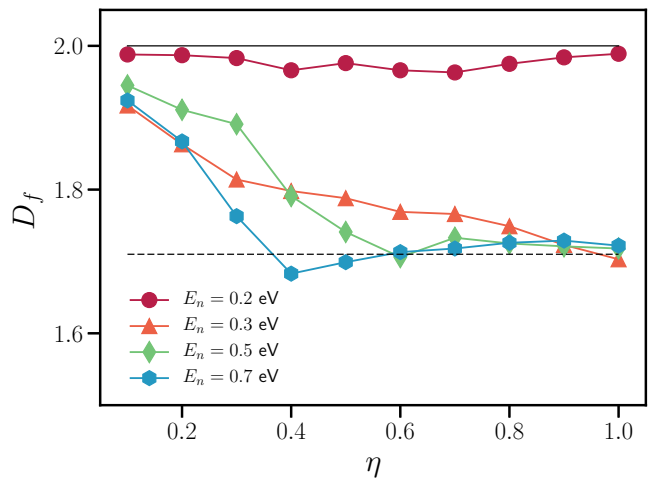}%
	\caption{\label{Average cluster shape} (Color online) Fractal dimension $D_{f}$ of clusters at interaction energies $E_{n}$ as function of interaction anisotropy $\eta$. The solid line corresponds to the dimension of compact two-dimensional objects ($D_{f} = 2$), while the dotted line represents $D_{f}$ of DLA clusters in 2D ($D_{f} = 1.71$).}
\end{figure}

We observe that strong interaction anisotropy ($\eta = 0.1$) leads to a visible elongation of clusters in x-direction (direction of strong in-plane bonds) for both considered values of $E_{n}$. Moreover, we find that clusters at $E_{n} = 0.7$ eV are stronger elongated and have a smaller width $W$ compared to clusters at $E_{n} = 0.2$ eV. This already suggests stronger impact of interaction anisotropy on cluster shapes at higher $E_{n}$. Further, strong interaction anisotropy ($\eta = 0.1$) removes the ramified structure of clusters at $E_{n} = 0.7$ eV, yielding elongated but compact clusters with smooth boundaries. This is confirmed in Fig. 3, where, upon decreasing $\eta$ from $1$ to $\eta = 0.1$, $D_{f}$ (at $E_{n} = 0.7$ eV) first remains unaffected up to $\eta = 0.4$ and then steadily increases until it approaches $D_{f} \approx 2$ for $\eta \rightarrow 0$. Later on (in Sec. III, C) we give an explanation for the interaction energy-dependent value of $\eta$ (in the regime $E_{n} \ge 0.3$ eV) where $D_{f}$ starts to increase from $D_{f} \approx 1.7$ upon decrease of $\eta$.

The emergence of elongated clusters reveals an imbalance between attachment and detachment rates for in-plane bonds in x- and y-direction, respectively. This is understandable from the fact that, for $\eta < 1$, the hopping rate $r_{ij}$ [see Eq. $(1)$] of a particle with in-plane bonds in y-direction only is higher compared to particles with in-plane bonds in x-direction because at $\eta < 1$ the inequality $\eta E_{n} = E_{n}^{y} < E_{n}^{x}$ holds. 

To summarize these observations, as $\eta$ is decreased from $1$, clusters become elongated for any value of the interaction energy $E_{n}$. At sufficiently large interaction energy ($E_{n} \le 0.3$ eV at $F = 1$ ML/min), one obtains DLA clusters at isotropic growth conditions. At low interaction energy ($E_{n} \le 0.2$ eV), $D_{f} \approx 2$ holds at any value of $\eta$. Different from that, at high interaction energy ($E_{n} \ge 0.3$ eV), the fractal dimension $D_{f}$ reveals a pronounced increase from the value $D_{f} \approx 1.7$ corresponding to the DLA universality class to a value ($D_{f} \approx 2$) reflecting regular, compact two-dimensional objects, as $\eta \rightarrow 0$.

So far we have concentrated on the directly visible differences in the spatial configurations of individual clusters at low and high interaction energy at isotropic and strongly anisotropic interactions. One also observes an impact of interaction anisotropy ($\eta < 1$) on the distribution of cluster sizes $S$. This is shown for the normalized cluster size distribution $P(S)$ in Fig. 4. For isotropic interactions and $E_{\text{n}} = 0.2$ eV, $S$ is rather equally distributed around a mean value of $S \approx 30$. This peak vanishes for interaction anisotropies $\eta < 0.4$ and $P(S)$ becomes flat in the region $S \ge 20$. Instead, $P(S)$ peaks at $S < 5$ for $\eta < 0.4$, which reflects a mixture of a few large and many very small clusters. Thus, the interaction anisotropy leads to a completely different composition of cluster sizes in the system. The reduced amount of large clusters is due to the lowered interaction energy $E_{n}^{y}$ of bonds in y-direction at $\eta < 1$. It follows that the detachment rate of particles with in-plane bonds in y-direction only increases as $\eta$ is decreased. Consequently, less stable clusters are formed and at $\theta = 0.05$ only a few clusters managed to surpass the critical cluster size to become stable. The situation is different at $E_{n} = 0.7$ eV. Decreasing the anisotropy parameter from $\eta = 1$ to $\eta = 0.4$ does not affect $P(S)$. Not until $\eta < 0.4$, the peak of $P(S)$ shifts from $S \approx 25$ to $S \approx 20$ which only reflects a marginal change in the cluster size distribution. Further, the distribution becomes narrower and clusters of size $S > 40$ vanish but the majority of clusters still has an intermediate size of $S \approx 20$ that is quite similar to the value at isotropic growth conditions.

\begin{figure}
	\includegraphics[width=1.0\linewidth]{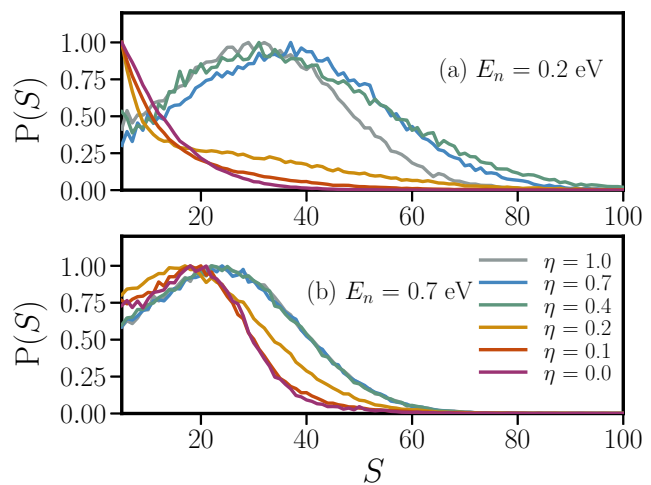}​
	\caption{\label{Average cluster shape} (Color online) Distribution $P(S)$ of cluster sizes $S$ at interaction energy $E_{n} = 0.2$ eV (a) and $E_{n} = 0.7$ eV (b) and various values of the anisotropy parameter $\eta$ ranging from isotropic conditions ($\eta = 1$) to $\eta = 0$ (no interaction energy for bonds along y-direction). Here and in the following figures, temperature $T$ and adsorption rate $F$ are chosen as in Fig. 2.}
\end{figure}

\begin{figure}
	\includegraphics[width=1.0\linewidth]{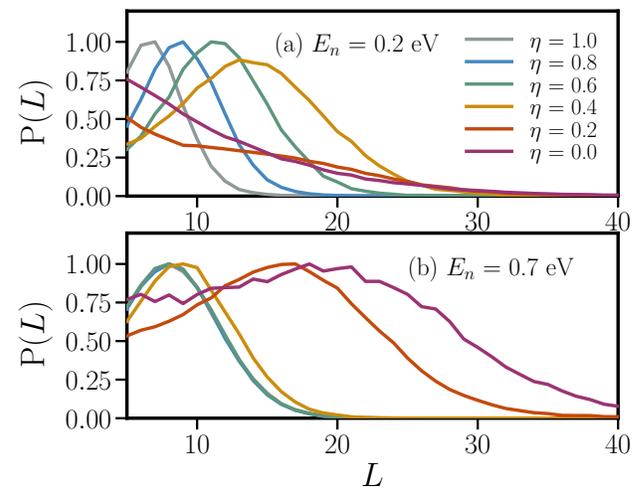}
	\caption{\label{Average cluster shape} (Color online) Distribution $P(L)$ of cluster lengths $L$ at $E_{n} = 0.2$ eV (a) and $E_{n} = 0.7$ eV (b) and various values of the anisotropy parameter $\eta$.}
\end{figure}

Additionally, the normalized cluster length distribution $P(L)$ is shown in Fig. 5. At $E_{n} = 0.2$ eV, the peak in the distribution $P(L)$ is shifted to larger values of $L$ and is slightly broadening as $\eta$ is decreased from $1$. This reflects the cluster elongation process, which continues up to $\eta = 0.4$. For stronger interaction anisotropies we only observe a sharp peak at very small length $L$, consistent with the results in Fig. 4. At $E_{n} = 0.7$ eV, the length distribution is not affected when $\eta$ is decreased from $\eta = 1$ to $\eta = 0.4$. For $\eta < 0.4$ the peak in $P(L)$ is shifted towards larger lengths $L$. At the same time $P(L)$ broadens.

\begin{figure}
	\includegraphics[width=1.0\linewidth]{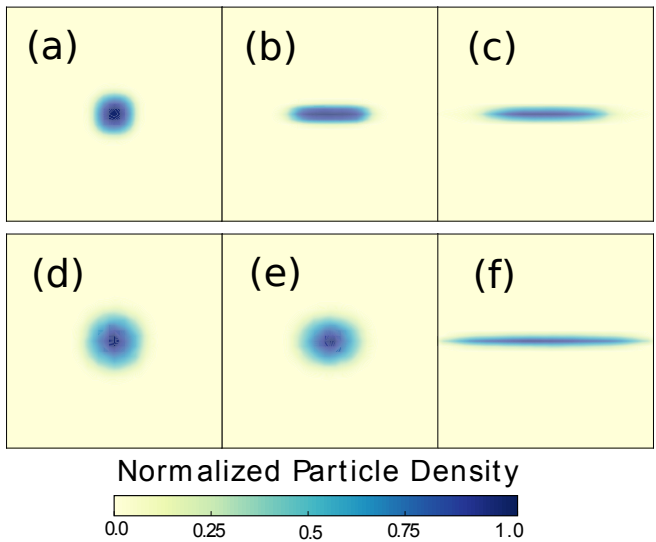}
	\caption{\label{Average cluster shape} (Color online) Average shapes of clusters of size $S = 40$ for different values of the interaction anisotropy parameter $\eta$ and the interaction energy $E_{n}$. The two values for $E_{n}$ considered are $E_{\text{n}} = 0.2$ eV (a-c) and $E_{n} = 0.7$ eV (d-f). The parameter $\eta$ is reduced from from isotropic interactions ($\eta = 1$) to $0.7$, $0.4$ and $0.1$.}
\end{figure}

\subsubsection{Average cluster shapes}

We now focus in more detail on the response of the average cluster shapes upon variations of the anisotropy parameter $\eta$. To this end we present in Fig. 6 the average cluster shapes for fixed cluster size $S = 40$ at $E_{\text{n}} = 0.2$ [see Fig. 6 (a)-(c)] and $E_{\text{n}} = 0.7$ eV [see Fig. 6 (d)-(f)] for three different values of the anisotropy parameter $\eta$. For $\eta = 1$, the average cluster shape is isotropic at both interaction energies, $E_{\text{n}} = 0.2$ eV and $E_{\text{n}} = 0.7$ eV, as expected. By decreasing $\eta$ from $1$ at $E_{n} = 0.2$ eV, the average clusters become immediately elongated along the direction of stronger interaction energy, i.e. along the x-direction. Thus, already relatively weak interaction anisotropy ($\eta = 0.7$) leads to anisotropic cluster shapes with growth preferred in x-direction. We conclude that at $E_{n} = 0.2$ eV there is a gradual cluster shape transformation as function of $\eta$ as soon as the regime of anisotropic interactions is entered ($\eta < 1$). For strong interaction anisotropy, such as $\eta = 0.1$, we observe strongly elongated clusters whose average shape [see Fig. 6(c)] matches quite good with the individual clusters shown in Fig. 2(b).

At $E_{n} = 0.7$ eV we encounter a different behavior of the cluster shape transformation. First, decreasing $\eta$ from $1$ to $\eta = 0.4$ has essentially no impact on the initially isotropic shape. Second, at $\eta = 0.1$, the clusters are much stronger elongated compared to the case $E_{n} = 0.2$ eV. This is in good agreement with the spatial configurations shown in Fig. 2(b) and (d). We conclude that there are two types of the cluster shape transformation, that is, gradual ($E_{n} = 0.2$ eV) versus sharp ($E_{n} = 0.7$ eV).

\begin{figure}
	\includegraphics[width=1.0\linewidth]{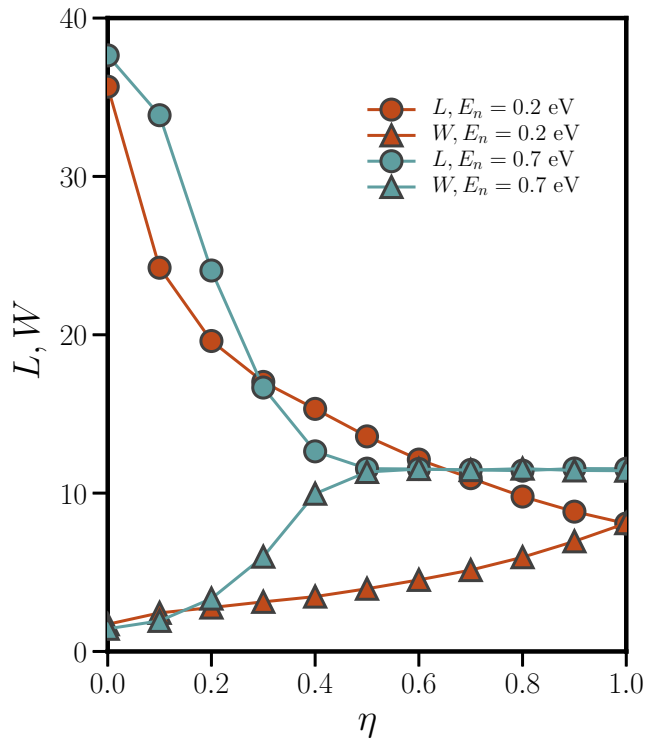}%
	\caption{\label{Average cluster shape} (Color online) Evolution of cluster length $L$ and width $W$ at $E_{n} = 0.2$ eV and $E_{n} = 0.7$ eV as function of $\eta$. The considered clusters are of size $S = 40$. The evolution of $L$ and $W$ upon decrease of $\eta$ from $1$ confirms the gradual shape transformation at $E_{n} = 0.2$ eV and the sharp shape transition at $E_{n} = 0.7$ eV.}
\end{figure}

To further illustrate that the type of the cluster shape transformation depends on the interaction energy $E_{n}$, the spatial extension of clusters (of size $S = 40$), namely the cluster length $L$ and width $W$ upon decreasing $\eta$ from $1$ are plotted in Fig. 7. At $E_{\text{n}} = 0.2$ eV we observe an immediate splitting of $L$ and $W$. This corresponds to anisotropic cluster growth where the cluster length and width grow at different rates, and thus, leads to the formation of elongated clusters. The smooth behavior of the splitting of $L$ and $W$ as function of $\eta$ at $E_{n} = 0.2$ eV confirms a gradual cluster shape transformation. In contrast, decreasing $\eta$ from $1$ at $E_{n} = 0.7$ eV leaves $L$ and $W$ essentially identical up to $\eta = 0.4$. Only for $\eta < 0.4$ we notice the splitting, which, in agreement with the results in Fig. 2 and Fig. 6, is also much stronger pronounced compared to $E_{n} = 0.2$ eV and therefore the cluster shape transformation is sharp.

Finally, it is interesting to discuss these results from the perspective of a simple model based on energy minimization. Indeed, at least close to equilibrium, one would expect that the cluster shapes are simply related to the total energy cost $E_{b}$ to form a compact cluster. For a cluster of size $S = L W$ with length $L$ and width $W$, $E_{b}$ is determined by the total number of cluster boundary sites and the corresponding energy of all broken bonds,
	
\begin{equation}
E_{b} = 2 W E_{n} + 2 L \eta E_{n}.
\end{equation}	

For a perfectly compact cluster of size $S$, the width is given by $W = S / L$. This allows us to rewrite Eq. (13) as

\begin{equation}
E_{b} = 2 E_{n} (S / L + \eta L),
\end{equation}
	
From Eq. (14) we find that $E_{b}$ reaches the minimum value when the cluster length fulfills 
	
\begin{equation}
L(S,\eta) = \sqrt{S / \eta}.
\end{equation}		
	
Equation (15) provides an estimate for the cluster length evolution in presence of anisotropic interactions at growth conditions, where particle detachment ($i^{*} > 1$, see Sec. III, C) is possible, such that clusters can obtain the equilibrium shape. We find that Eq. (15) indeed describes the cluster length at small adsorption rates $F$ ($F = 1$ ML/min in Fig. 8) and low interaction energies $E_{n}$ (i.e., under conditions where particle detachment is indeed present). This is shown in Fig. 8 where, as function of $\eta$, the cluster length $L_{40}$ (of clusters of size $S = 40$) is plotted for different $E_{n}$ at $F = 1$ ML/min. Consider, as an example, the value $E_{n} = 0.15$ eV. Here, the rate for breaking lateral bonds (of particles with one nearest-neighbor) is high enough in both, the x- and y-direction, such that clusters retain a compact shape (the interaction energies $E_{n}$ where attachment is either reversible or irreversible are discussed in detail in Sec. III C). Consequently, $L_{40}$ is close to the length predicted by Eq. (15) in the range $0.1 \le \eta \le 1.0$ (see the black line in Fig. 8). Finally, at $E_{n} = 0.3$ eV, clusters (at $\eta = 1$) are ramified with fractal dimension $D_{f} \approx 1.7$ (see Fig. 3), which means that they are far from the compact equilibrium shape. As a consequence, we observe large deviations from the equilibrium length predicted by Eq. (15) also in the regime of anisotropic interactions (this holds for any $E_{n} \le 0.3$ eV). Taken together, these results reflect the fact that energetic arguments expressed by Eqs. (13)-(15) only hold at small values of $E_{n}$ and $F$, where particle detachment is possible ($i^{*} > 1$) for any $\eta$.

\begin{figure}
	\includegraphics[width=1.0\linewidth]{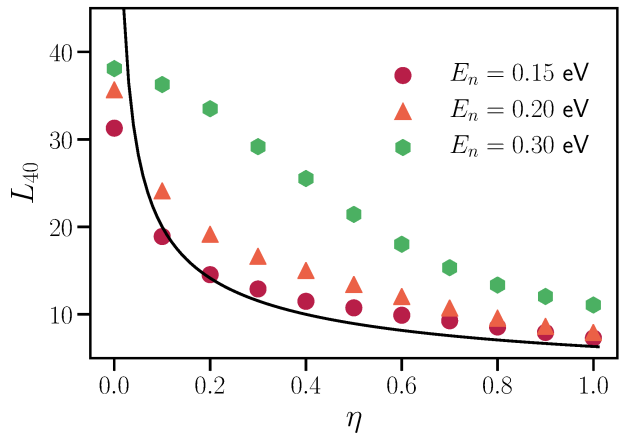}%
	\caption{\label{Average cluster shape} (Color online) Length $L_{40}$ of clusters of size $S = 40$ as function of the interaction anisotropy parameter $\eta$ for different interaction energies $E_{n}$ at $F = 1$ ML/min. The solid black line corresponds to the length $L(S,\eta) = \sqrt{S / \eta}$ for growth close to equilibrium [see Eq. (15)].}
\end{figure}

\subsection{Cluster shape properties - One-dimensional vs. two-dimensional cluster growth}

\subsubsection{Cluster length evolution}

The results presented so far already demonstrate the impact of interaction anisotropy on shape properties of clusters under non-equilibrium growth conditions. Now we focus on the evolution of cluster shapes during growth. To this end we plot in Fig. 9 the average cluster length $L(S)$ as function of cluster size $S$ at $E_{n} = 0.2$ eV and $E_{n} = 0.7$ eV for different values of $\eta$. At $E_{n} = 0.2$ eV and isotropic interactions ($\eta = 1$), $L(S)$ closely follows the prediction from energy arguments [see Eq. (15)], i.e. $L = \sqrt{S}$. In contrast, at $E_{n} = 0.7$ eV and $\eta = 1$ we observe significant deviations because the clusters are now ramified and do not exhibit the equilibrium shape [see Fig. 2(c)]. 

\begin{figure}
	\includegraphics[width=1.0\linewidth]{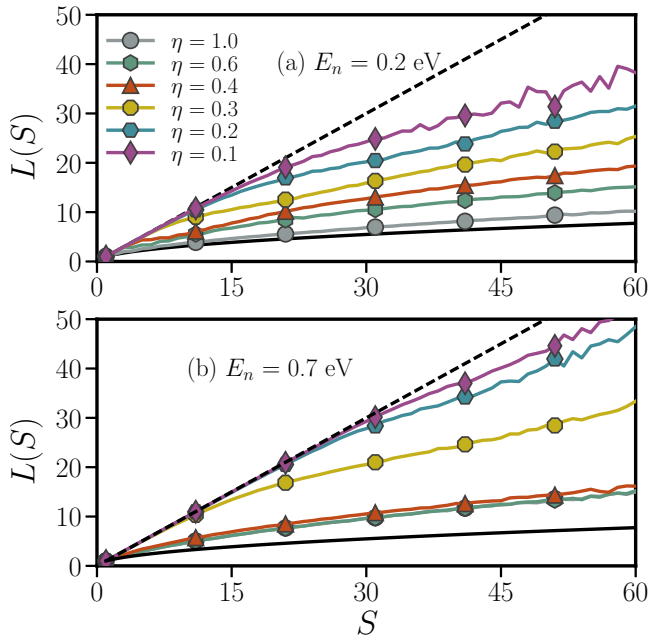}%
	\caption{\label{Average cluster shape} (Color online) Average cluster length $L(S)$ as function of cluster size $S$ for interaction energy $E_{n} = 0.2$ eV (a) and $E_{n} = 0.7$ eV (b) and different values of $\eta$. The solid line represents $L(S) = \sqrt{S}$, while the dotted line corresponds to $L(S) = S$.}
\end{figure}

As interaction anisotropy is switched on ($\eta < 1$), we observe an immediate effect on the evolution of $L(S)$ at $E_{n} = 0.2$ eV (i.e., at relatively weak interaction anisotropy). This is consistent with the previous results and confirms a gradual cluster shape transformation. In contrast, at $E_{n} = 0.7$ eV, $L(S)$ remains unchanged up to $\eta \ge 0.4$, (see the snapshots in Fig. 6). This finding approves again that at high interaction energy ($E_{\text{n}} = 0.7$ eV), weak interaction anisotropy ($\eta \ge 0.4$) has far less impact on the cluster shape than at low interaction energies.

As $\eta$ is lowered, one finds a transition to a linear relation $L(S) = S$ (see the dotted line in Fig. 9) for small cluster sizes $S$. The relation $L(S) = S$ represents maximally elongated, one-dimensional clusters which grow in the direction of the strong bonds only (''one-dimensional growth''). At $E_{n} = 0.7$ eV and $\eta = 0.1$, $L(S)$ follows the one-dimensional growth line up to $S \approx 30$, while at $E_{n} = 0.2$ eV and $\eta = 0.1$, $L(S)$ begins to deviate from $L(S) = S$ already at around $S \approx 10$. Therefore, the one-dimensional growth is more robust at high interaction energies.

We conclude that in the presence of strongly anisotropic interactions, the initial stage of cluster growth appears to be one-dimensional with respect to particle attachment. This growth mode breaks down at a specific cluster length $L_{c}$ which depends on $E_{n}$ and $\eta$. The length $L_{c}$, is defined as the length of the cluster where $|L(S)-S| \ge 1$ sets in upon increase of $S$. The value of $L_{c}$ increases with increasing $E_{n}$, resulting in stronger elongated clusters at high interaction energies $E_{n}$. 

\subsubsection{The breakdown of one-dimensional cluster growth}

In Fig. 10, $L_{c}(\eta)$ is plotted for various interaction energies as function of $\eta$. Irrespective of $E_{n}$, the function $L_{c}(\eta)$ increases as $\eta$ is decreased from $1$ and converges to a finite value in the limit $\eta \rightarrow 0$. Specifically, in the range $E_{n} \ge 0.3$ eV, $L_{c}$ converges to very similar values $L_{c}(0) = L_{c}^{0} \approx 35$. In contrast, at $E_{n} < 0.3$ eV, $L_{c}^{0}$ depends on $E_{n}$, e.g. $L_{c}^{0} \approx 25$ at $E_{n} = 0.2$ eV and $L_{c}^{0} \approx 18$ at $E_{n} = 0.15$ eV. We also see that, at $E_{n} \le 0.3$ eV, the increase of $L_{c}$ as function of $\eta$ is smooth and sets in already at weak interaction anisotropy. In contrast, at $E_{n} > 0.3$ eV, $L_{c}$ remains essentially unaffected by weak interaction anisotropy. Also shown in Fig. 10 is the function $L_{c}(\eta)$ resulting from energy considerations (derived from Eq. (15)). We observe that this function yields a reliable estimate only for small values of $E_{n}$ (consistent with the discussion of $L_{40}$ in Fig. 8).

\begin{figure}
	\includegraphics[width=1.0\linewidth]{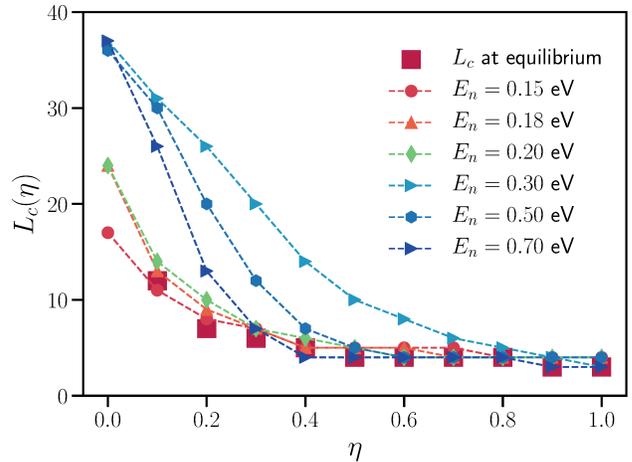}​
	\caption{\label{Average cluster shape} (Color online) Critical length $L_{c}(\eta)$ for various values of the interaction energy $E_{\text{n}}$ as function of the anisotropy parameter $\eta$. For comparison, we have included the corresponding result from energy considerations [see Eq. (15)].}
\end{figure}

\subsubsection{Transition to two-dimensional self-similar cluster growth }

To better characterize the growth mode that follows the one-dimensional growth upon increase of the cluster size $S$, we calculate the average aspect ratio $R(S) = L(S)/W(S)$, where $W(S)$ is the cluster width. An aspect ratio $R$ that remains constant as function of cluster size $S$ implies that the cluster growth is self-similar. This means, in other words, that $L(S)$ and $W(S)$ increase at constant rates. Fig. 11 (a) shows $R(S)$ for $\eta \le 0.4$ and different values of $E_{n}$. After an initial linear increase, corresponding to the region of one-dimensional cluster growth, the aspect ratio $R(S)$ reaches a plateau and then remains constant as $S$ increases. We call this saturation value $R^{sat}$. A saturation value $R^{sat} = 1$ would correspond to isotropic self-similar growth without preferred growth direction [i.e., $L(S) = W(S)$]. For $R^{sat} > 1$, clusters are elongated [$L(S) > W(S)$] but the growth is still self-similar. Such a plateau exists for all considered combinations of $E_{n}$ and $\eta$. Only the actual value of $R^{sat}$ and the cluster size $S$, where the plateau is reached, depend specifically on $E_{n}$ and $\eta$. 

In addition, $R^{sat}$ is plotted for various $E_{n}$ as function of $\eta$ in Fig. 11 (b). For $E_{n} \ge 0.3$ eV, $R^{sat}$ converges to similar values as $\eta \rightarrow 0$. The onset of anisotropic self-similar growth ($R^{sat} > 1$) is shifted to smaller values of $\eta$ for increasing $E_{n}$. Furthermore, for $E_{n} < 0.3$ eV, $R^{sat}$ is lower compared to $R^{sat} \approx 25$ as for $E_{n} \ge 0.3$ eV.

\begin{figure}
	\includegraphics[width=1.0\linewidth]{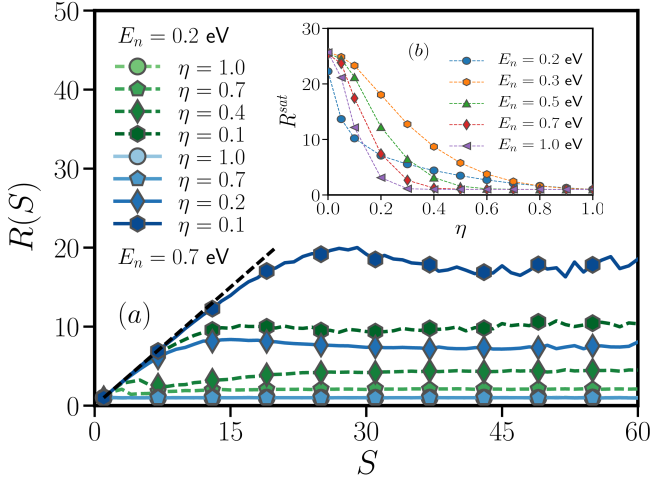}%
	\caption{\label{Average cluster shape} (Color online)  (a) Evolution of the average aspect ratio $R(S) = L(S)/W(S)$ as function of cluster size $S$ at $E_{\text{n}} = 0.2$ eV (greens) and $E_{\text{n}} = 0.7$ eV (blues) at different values of $\eta$. The black dotted line corresponds to $R(S) = S$. (b) Saturation value $R^{sat}(\eta)$ for various interaction energies $E_{n}$}
\end{figure}

\subsection{Role of the critical cluster size $i^{*}$}
	
A major observation in presence of anisotropic interactions ($\eta < 1$) is that, depending on the interaction energy $E_{n}$, there exist two types of cluster shape transformations. It turns out that this can be explained by properties of the isotropic reference systems ($\eta = 1$). To this end, we now	take a closer look at the critical cluster size $i^{*}$ at $\eta = 1$ as function of interaction energy $E_{n}$.

For this purpose, the critical cluster density $\rho_{N}^{c}$ is plotted as function of $E_{n}$ for the isotropically interacting reference systems ($\eta = 1$) in Fig. 12. We recall that $\rho_{N}^{c}$ scales $\sim (D_{0}(T)/F)^{-\chi}$ [with $\chi = i^{*} / (i^{*} + 2)$]. From Fig. 12 it is seen that $\rho_{N}^{c}$ increases as the interaction energy is increased from $E_{n} = 0.1$ eV to $E_{n} = 0.3$ eV, but saturates in the range $E_{n} > 0.3$ eV. Since we do not change the temperature $T$ and adsorption rate $F$, it follows from the known scaling of $\rho_{N}^{c}$ \cite{theo_rate_1, theo_rate_3, theo_rate_5, theo_rate_6} that $i^{*}$ has to be identical for all $E_{n} > 0.3$ eV. In this regime, the critical cluster size therefore is $i^{*} = 1$, which corresponds to conditions where bonds are $\textit{irreversible}$.

\begin{figure}
	\includegraphics[width=1.0\linewidth]{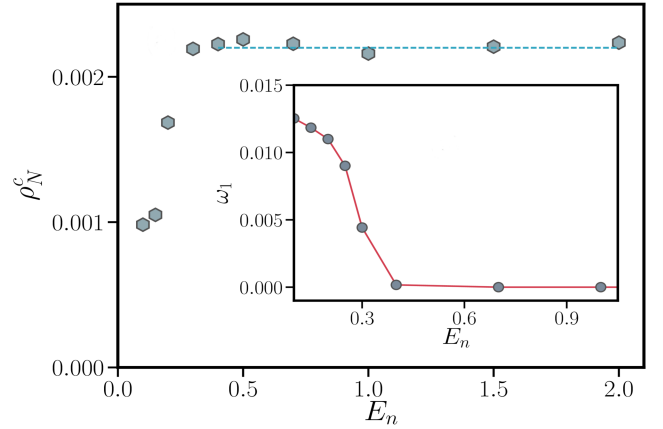}%
	\caption{\label{Average cluster shape} (Color online) Critical cluster density $\rho_{N}^{c}$ for different in-plane interaction energies $E_{n}$ in the (isotropic) reference system. The inset shows the detachment ratio $\omega_{1}$ as function of $E_{n}$.}
\end{figure}

This conclusion is confirmed by the analysis of $\omega_{1}(E_{n})$ [see Eq. (9)] at $\eta = 1$ (see the inset of Fig. 12). Consistent with the analysis so far, $\omega_{1} = 0$ for $E_{n} > 0.3$ eV, which means absence of particle detachment (bonds are $\textit{irreversible}$). In other words, already dimers form stable clusters and $i^{*} = 1$. As $E_{n}$ is decreased towards lower values, $\omega_{1}$ becomes nonzero, which means that bonds become \textit{reversible} (particles can break bonds and detach from clusters). Consequently, one enters the regime of $i^{*} > 1$. We conclude that, coming from high interaction energies, there is a transition from \textit{irreversible} to \textit{reversible} bonds at $E_{n} \approx 0.3$ eV. This observation allows to explain the type and the onset of the cluster shape transformation in presence of anisotropic interactions ($\eta < 1$). We can distinguish between three different scenarios.

\subsubsection{Interaction energy $E_{n} > 0.3$ eV and $\eta = 1$}

In this case, bonds are \textit{irreversible} in x- and y-direction and consequently, cluster growth is isotropic for any $E_{n} > 0.3$ eV (see inset of Fig. 11 at $\eta = 1$ where $R^{sat} = 1$ at $\eta = 1$). 

\subsubsection{Interaction energy $E_{n} > 0.3$ eV and $\eta < 1$}

In this situation, bonds in x-direction are \textit{irreversible} and as long as $E_{n}^{y} = \eta E_{n} > 0.3$, also bonds in y-direction are \textit{irreversible} and therefore, cluster growth is isotropic even for $\eta < 1$ (as long as $E_{n}^{y} > 0.3$ eV). However, as soon as the regime $E_{n}^{y} = \eta E_{n} < 0.3$ is entered by increasing the strength of the interaction anisotropy, bonds in y-direction become \textit{reversible} while bonds in x-direction remain \textit{irreversible}. Anisotropic cluster growth sets in at the value of $\eta$ that leads $E_{n}^{y} \le 0.3$ eV. At $E_{n} = 0.5$ eV, we find the onset of the transformation at $\eta = 0.6$, which corresponds to $E_{n}^{y} = 0.3$ eV, consistent with the transition from \textit{irreversible} to \textit{reversible} bonds. The same holds at $E_{n} = 1.0$ eV, where we find the onset at $\eta = 0.3$, which again corresponds to $E_{n}^{y} = 0.3$ eV for bonds along the y-direction. This explains, depending on $E_{n},$ the sharp cluster shape transformation and the precise value $\eta$ where it sets in.

\subsubsection{Interaction energy $E_{n} < 0.3$ eV and $\eta \le 1$}

Here, bonds in both, x- and y-direction are \textit{reversible}. At $\eta = 1$, cluster growth is isotropic because there is no imbalance between bond strengths, $E_{n}^{x} = E_{n}^{y}$. In contrast to case 2., clusters become elongated for any $\eta < 1$ because we are always in the regime of \textit{reversible} bonds ($E_{n}^{y} < 0.3$ for all $\eta$). Even though bonds in x-direction are \textit{reversible}, the detachment rate for bonds in y-direction is higher at $\eta < 1$. Therefore, clusters growth is anisotropic with growth preferred in x-direction (despite the fact that also bonds in x-direction are \textit{reversible}). Different from case $2.$, the cluster shape transformation here is gradual because we are always in the regime where bonds are \textit{reversible}. 
\\
\\
Moreover, we can also explain why both, $L_{c}$ and $R^{sat}$ converge to similar values for $E_{n} \ge 0.3$ as $\eta \rightarrow 0$ (see Fig. 10 and Fig. 11). In the latter case, $E_{n}^{y} = 0$. Therefore, all of the systems are identical in the sense that bonds in x-direction are $\textit{irreversible}$ while bonds in y-direction are \textit{reversible}. The hopping rate for particles with bonds in y-direction only is the same as the rate for free diffusion, $r_{ij} \sim \text{exp}\left( -\beta E_{d}\right)$. The reason for this is that at $\eta = 0$ their is no contribution to the activation energy barrier $\Delta E$ from in-plane bonds this direction. Consequently, $L_{c}(0)$ and $R^{sat}(0)$ is practically the same for $E_{n} \ge 0.3$ eV.

\subsection{Effect of the adsorption rate $F$ on cluster properties}

So far we have focused on the interplay between the interaction energy $E_{n}$ and the interaction anisotropy $\eta$ at fixed adsorption rate $F$ ($F = 1$ ML/min). In experimental studies, the parameters $E_{n}$ and $\eta$ are essentially fixed (corresponding to the system considered). The parameter, which can be precisely varied in experiments (next to the temperature $T$) is the adsorption rate $F$. This parameter has indeed a profound impact since it determines not only the cluster density, but also the thin film morphology in the multilayer growth regime (as it is well established for atomic growth \cite{Shen2004, Krug2000, Caflisch1999, Politi2001, Politi2002A, Politi2002B, Politi2003}). Motivated by this, we therefore study the impact of different adsorption rates $F$ in presence of anisotropic interactions. 
		
First, we consider the scaling of the average aspect ratio $\braket{R} = 1/N \sum R$ (sum of all aspect ratios $R$ in the system divided by $N$, the number of clusters) with adsorption rate $F$ in the regime of strong interaction anisotropy, $\eta \le 0.3$. Second, we focus on the evolution of the average cluster length $\braket{L}$ and width $\braket{W}$ as function of coverage $\theta$ before coalescence sets in.

Considering a coverage $\theta = 0.1$, we find that $\braket{R}$ exhibits power-law scaling, $\braket{R} \sim F^{-\alpha}$, with $\alpha \approx 0.3$. This is shown in Fig. 13 where $\braket{R}$ is plotted as function of $F$ at $E_{n} = 0.5$ eV and $\eta \le 0.3$ eV. Interestingly, the scaling exponent ($\alpha \approx 0.3$) does not depend on $\eta$ for all analyzed interaction energies from $E_{n} = 0.2$ eV to $E_{n} = 0.7$ eV.

\begin{figure}
	\includegraphics[width=1.0\linewidth]{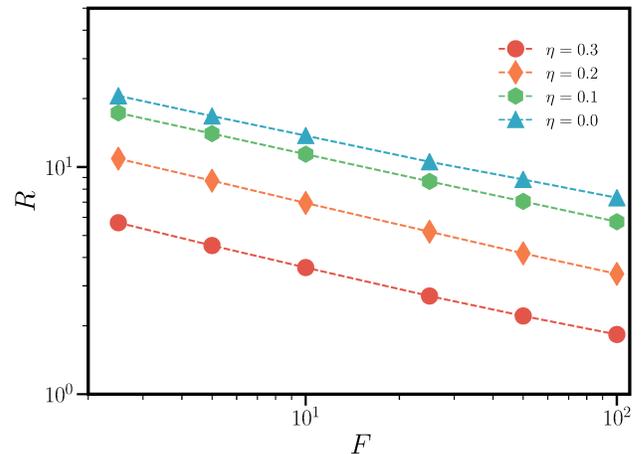}%
	\caption{\label{Average cluster shape} (Color online) Double logarithmic plot of the average aspect ratio $\braket{R}$ at $\theta = 0.1$ and $E_{n} = 0.5$ eV as function of adsorption rate $F$ in the regime of strong interaction anisotropy.}
\end{figure}	

\begin{figure}
	\includegraphics[width=1.0\linewidth]{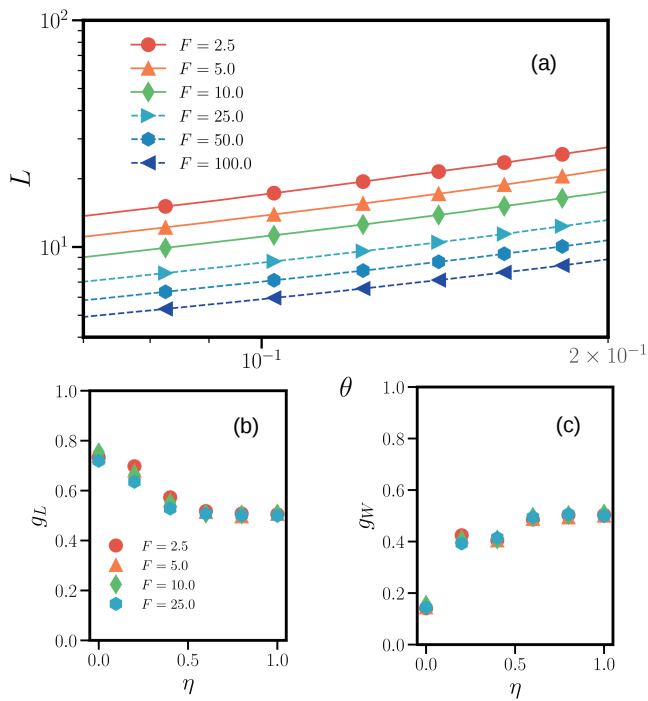}%
	\caption{\label{Average cluster shape} (Color online) (a) Double logarithmic plot of the average cluster length $\braket{L}$ as function of $\theta$ for various adsorption rates $F$ at $E_{n} = 0.5$ eV and $\eta = 1$. (b) Cluster length scaling exponent $g_{l}$ and (c) cluster width scaling exponent $g_{w}$ as function of $\eta$.}
\end{figure}

The evolution of the average length $\braket{L}$ versus coverage $\theta$ at $E_{n} = 0.5$ eV and $\eta = 0.3$ is plotted in Fig. 14 (a), where we have included results for different values of $F$. We observe again power-law scaling with scaling exponent $g_{l}$ that depends only weakly on the adsorption rate $F$ (we checked that the power-law scaling holds for any $E_{n}$ in the range from $E_{n} = 0.2$ eV to $E_{n} = 0.7$ eV). This is confirmed by Fig. 14 (b) where one can see that $g_{l}$ somewhat increases with increasing $\eta$, but remains almost identical for different adsorption rates $F$. Different from the behavior of $g_{l}$,  the exponent $g_{w}$ decreases for decreasing $\eta$ (again almost independent of $F$) as shown in Fig. 14 (c). Moreover, we find that, as long as the cluster growth is isotropic, the scaling exponents have values $g_{l} \approx g_{w} \approx 0.5$, which are consistent with those observed during the domain growth in the random-field Ising model with isotropic interactions (RFIM-DI) \cite{Bupathy18} or the Axial Next-Nearest-Neighbor Ising Model (ANNNI) \cite{Cheon2001}. When anisotropy is switched on in the RFIM-DI and the ANNNI, the scaling exponents become different along the x- and y-direction, respectively. In the present model, we also find different exponents for the two lattice directions, that is, $0 < g_{w} < 0.5 < g_{l} < 1$, when the interactions become strongly anisotropic. This indicates an interesting similarity between our growth model and the models of Ising type. 

\subsection{Comparison with the anisotropic stochastic Eden growth model}

So far we have investigated the cluster growth by event-driven kMC simulations where adsorption, nucleation, attachment, detachment and diffusion processes of particles on the substrate are included. We now turn to the much simpler anisotropic Eden model for cluster growth (see Sec. II, B), where anisotropy of lateral bonds is taken into account by an imbalance of attachment probabilities as described in Eq. (6). One main difference to the kMC model is that diffusion processes are absent in the Eden model, where the cluster growth is only determined by the attachment probabilities of boundary sites. Our key question is whether this minimal model still contains sufficient information to reproduce the clusters obtained in the kMC simulations for various anisotropic growth conditions. 

For this purpose, we plot in Fig. 15 the cluster length $L(S)$ obtained from both, kMC simulations at $E_{n} = 0.2$ eV and the Eden model. This value of $E_{n}$ is chosen because it produces compact clusters with fractal dimension $D_{f} \approx 2$. Moreover (as confirmed in Fig. 8), the cluster shape at $E_{n} = 0.2$ eV is close to the equilibrium shape for any $\eta$, and the Eden model with anisotropic interactions is supposed to correctly describe equilibrium clusters. The good agreement in Fig. 15 for the isotropic case ($\eta = \xi = 1$) is expected since there is no imbalance between attachment and detachment rates for the x- and y-direction. More interesting is the matching of both models in the regime where anisotropic interactions are present. Here, the anisotropy parameter $\xi$ for the attachment probabilities in the Eden model [see Eq. (6)] has been used as a fitting parameter to reproduce the kMC results. The Eden model does not only correctly describe the cluster length evolution $L(S)$, but also gives the correct critical lengths $L_{c}$ and the aspect ratio $R^{sat}$ in the self-similar growth regime. This is shown in Fig. 16 (a) and Fig. 16 (b), respectively. We note, however, that the anisotropic Eden model is not able to reproduce clusters at interaction energies $E_{n}$ where cluster shapes deviate from the equilibrium shape (see $L_{40}$ at $E_{n} = 0.3$ eV in Fig. 8). For adsorption rate $F = 1$ ML/min, this means that the Eden model is only capable to reproduce clusters at $E_{n} \le 0.2$ eV.

The good agreement between the two approaches (for $E_{n} \le 0.2$ eV in the kMC simulations at $F = 1$ ML/min) shows that the isotropic diffusion of free particles does not play a crucial role in the cluster formation process in presence of anisotropic interactions. However, this is only the case as long as one operates in the regime where detachment of particles from clusters is possible ($i^{*} > 1$). In this regime, it seems that only the rates for attachment and detachment determine the cluster shape. Therefore, we expect that a more detailed analysis of attachment and detachment rates will generally (also for the case $i^{*} = 1$) help to better understand the resulting cluster shapes under non-equilibrium growth conditions in presence of anisotropic interactions.

\begin{figure}
	\includegraphics[width=1.0\linewidth]{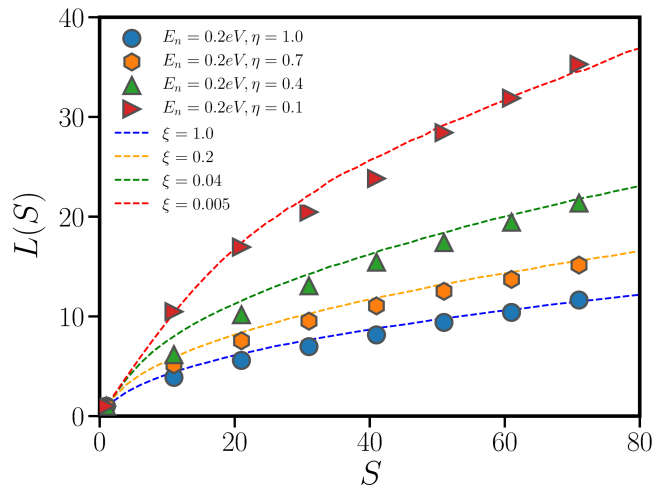}%
	\caption{\label{Average cluster shape} (Color online) Comparison of the cluster length evolution between clusters obtained from kinetic Monte-Carlo simulations (at $E_{\text{n}} = 0.2$ eV and different values of $\eta$) with clusters obtained from the stochastic Eden model at different ratios $\xi$ of the attachment probability in x- and y-direction, respectively.}
\end{figure}

\begin{figure}
	\includegraphics[width=1.0\linewidth]{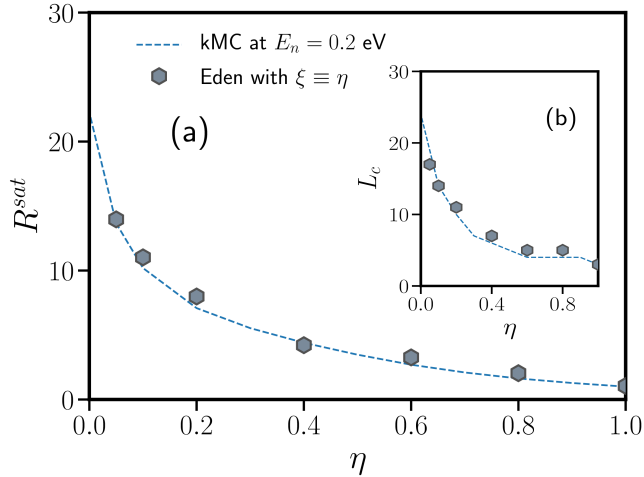}%
	\caption{\label{Average cluster shape} (Color online) (a) Saturation value of the aspect ratio $R^{sat}$ and (b) the critical length $L_{c}$ of clusters obtained from kinetic Monte-Carlo simulations and the Eden model. The in-plane interaction energy in kMC simulations is $E_{\text{n}} = 0.2$ eV. The anisotropy parameter $\eta \in [0,1]$ and $\xi$ is modified such that it fits the kMC results.}
\end{figure}

\section{Conclusions}

Using event-driven kMC simulation on a square-lattice we have studied the effect of anisotropic nearest-neighbor interactions in the sub-monolayer growth regime. Our model assumes a spherical particle shape and anisotropy is introduced by reducing the interaction energy of in-plane bonds along the y-direction by a factor $\eta$ relative to the interaction energy in x-direction. By varying the interaction energy, the anisotropy parameter and the adsorption rate, we have analyzed in detail the resulting clusters in presence of anisotropic interactions.

As expected, anisotropic interactions lead to non-spherical (elongated), rod-like and needle-shaped clusters with fractal dimension $D_{f} \approx 2$, as we have shown explicitly by inspecting snapshots and by calculating the fractal dimension as function of interaction anisotropy. Moreover, we have shown that energetic arguments for cluster shapes only hold for low adsorption rates and low interaction energies. Furthermore, cluster size distributions show that, for increasing interaction anisotropy, clusters become smaller. This effect is the more pronounced the smaller interaction energy.

A detailed analysis of cluster shapes as function of $\eta$ reveals two different types of cluster shape transformation. At low interaction energy, the transformation from isotropic to elongated clusters is gradual. In contrast, it is sharp at high interaction energies. Moreover, for strong interaction anisotropy, the early stage of cluster growth appears to be one-dimensional with particle attachment along the direction of strong bonds only. This growth mode breaks down at a critical length. From analyzing the aspect ratio we identify the subsequent self-similar growth mode.

Interestingly, we have found that the critical cluster density and detachment ratio in the isotropic reference system help to explain the properties of the cluster shape transformation in the anisotropic case. Furthermore, the comparison with the reference systems also explains the value of the anisotropy parameter where the transformation from isotropic to elongated cluster shapes sets in.

Moreover, we have investigated the effect of the (experimentally controllable) adsorption rate on cluster shape properties. According to our results, the average aspect ratio as function of adsorption rate displays power law scaling in the regime of strong interaction anisotropy. The scaling exponent does not depend on the anisotropy parameter. Also, the evolution of the average cluster length and width as function of coverage exhibit power-law scaling with scaling exponents that depend only weakly on the adsorption rate.

In addition to kMC simulations, we have also employed an anisotropic version of the Eden model where diffusion processes are neglected. In this context, we have used the anisotropy parameter, that controls the attachment probabilities, as a fitting parameter. By this it is indeed possible to reproduce main features of the cluster growth observed in the kMC simulations. In particular, we find good agreement in the cluster length evolution, the critical length and the saturation value of the aspect ratio. 

The good agreement between results from kMC simulations and the Eden model suggests that attachment (rather than diffusion) is the dominant mechanism in determining cluster shapes. Therefore it may be worth to further investigate, on a very fundamental level, attachment as well as detachment rates to gain a deeper understanding of the cluster shape transformations. 

The present kMC simulations can be extended in several directions. By appropriate setting the values for the anisotropy parameter in the kMC setup, our model could be used to numerically study the experimentally relevant growth of elongated Zn clusters on isotropic surfaces \cite{Lu2016} as well as to study the effect of dipole-dipole interactions among particles. By this one could approach experimentally relevant systems that have already been studied \cite{DellaSala2011, exp_xray_3}. A further future direction would be to explore, based on our model, the multilayer growth regime. Investigations in these directions are in progress.

\begin{acknowledgments}
This  work  was  supported  by  the  Deutsche  Forschungsgemeinschaft  within
the framework of the Collaborative Research Center CRC 951 (project A7). We also thank J. Dzubiella and M. Miletic for fruitful discussions. 
\end{acknowledgments}

\bibliography{basename of .bib file}

\end{document}